\documentclass[a4paper,fleqn]{cas-dc}
\renewcommand{\printorcid}{}
\usepackage{natbib}

\usepackage[T1]{fontenc}
\usepackage{ae,aecompl}
\usepackage{graphicx}
\usepackage{epstopdf}
\usepackage{verbatim}
\usepackage{amsmath}	
\usepackage{amssymb}	
\usepackage{tikz}
\usepackage{float}
\usepackage{footnote}
\usepackage{amssymb}
\usepackage{ulem}
\usepackage{hyperref}
\hypersetup{
    colorlinks=true,
    linkcolor=teal,
    citecolor=teal,
    urlcolor=teal
}

\definecolor{grey}{rgb}{0.5,0.6,0.7}

\definecolor{amber}{rgb}{1.0,0.49,0.0}

\newcommand{\Z}{{\it Z} }
\newcommand{\gr}{$g-r$ }
\newcommand{\smass}{$M^*$ }
\newcommand{\hb }{H$\beta$}
\newcommand{\fe }{$\langle\rm{Fe}\rangle$}
\newcommand{\smassa}{$M^*$}

\begin{document}
\let\WriteBookmarks\relax
\def\floatpagepagefraction{1}
\def\textpagefraction{.001}
\shorttitle{Age and Metallicity of galaxies in the Coma Supercluster}
\shortauthors{Tiwari et~al.}

\title [mode = title]{Age and Metallicity of galaxies in different environments of the Coma supercluster}

\author[1]{Juhi Tiwari}
\cormark[1]
\ead{tiwarijuhi92@gmail.com}
\author[1]{Smriti Mahajan}
\cormark[2]
\ead{mahajan.smriti@gmail.com}
\author[1]{Kulinder Pal Singh}
\address[1]{Department of Physical Sciences, Indian Institute of Science Education and Research Mohali, Knowledge City, Sector 81, Manauli 140306, Punjab, India}
\cortext[cor1]{Corresponding author}
\cortext[cor2]{Principal corresponding author}

\begin{abstract}
We analyse luminosity-weighted ages and metallicity (\Z) of galaxies in a continuous range of environments, i.e. clusters, filaments and voids prevalent in the Coma supercluster ($\sim 100 h^{-1}$ Mpc).  
 Specifically, we employ two absorption line indices, H$\beta$ and \fe~as tracers of age and metallicity of galaxies. We find that the stellar-phase metallicity of galaxies declines with increasing age as 
 a function of stellar mass ($M^*$) as well as environment. On the filaments, metallicity of galaxies varies as a function of their distance from the spine of the filament, such that galaxies closer to the centre of 
 the filaments have lower metallicity relative to their counterparts 1 Mpc away from it. The mean age of intermediate mass galaxies ($10^{10} < M^*/M_{\odot} < 10^{10.5}$) galaxies is statistically 
 significantly different in different 
 environments such that, the galaxies in clusters are older than the filament galaxies by 1-1.5 Gyr, while their counterparts in the voids are younger than filament galaxies by $\sim 1$ Gyr. The massive 
 galaxies ($M^*/M_{\odot} > 10^{10.5}$), on the other hand show no such difference for the galaxies in clusters and filaments, but their counterparts in voids are found to be younger 
 by $\sim 0.5$ Gyr. At fixed age however, \Z of galaxies is independent of their \smass in all environments, except the most massive ($M^*/M_{\odot} \gtrsim 10^{10.7}$), oldest galaxies 
 ($\gtrsim 9$ Gyr) which show a sharp decline in their \Z with \smassa. Our results support a scenario where galaxies in the nearby Universe have grown by accreting smaller galaxies or primordial gas from the large-scale cosmic web. 
\end{abstract}
\begin{NoHyper}
\maketitle
\end{NoHyper}
\section{Introduction}
 Metallicity ({\it Z}) is a fundamental attribute which provides clues to the star formation history of a galaxy. Just like other global properties such as colour, \Z is also strongly correlated with the stellar mass of galaxy \citep{lequeux79}. The observed trend is such that the \Z increases linearly with stellar mass \citep{caldwell03,ogando08,trager08} or luminosity \citep{skillman89,brodie91,bell00} at low masses, but saturates for massive galaxies ($M^*/M_{\odot} \gtrsim 10^{10.5}$). Literature  \citep{panter08,leethochawalit18} on the evolution history of \Z through cosmic time shows that this trend holds well for stellar \citep[e.g.][]{gallazzi05} as well as gas phase metallicity \citep[e.g.][]{tremonti04,cooper08} in galaxies. The observed trend is also well established for galaxies at low \citep[$z\sim0.1$; e.g.][]{gallazzi05} and high redshift \citep{erb06} alike. 
 
 The analysis of metallicity in galaxies is, however, complicated by the fact that observables such as broadband colours and spectral line indices which are sensitive to chemical enrichment, also respond in a similar way to changes in luminosity-weighted age. For instance, colours of the dominant stellar population become redder not only with increasing age but also with stellar {\it Z}. The variation due to age occurs because more stars move to the giant branch, while an increase in the {\it Z} leads to an increase in the stellar opacity. In such a situation the dust in the interstellar medium (ISM) re-emits the blue photons absorbed from stars at longer wavelengths, making a galaxy appear red. Since the optical colours may redden due to age or {\it Z}, or both, two galaxies with different age and chemical enrichment may exhibit similar colours and spectral features of comparable strengths. This is famously known as the age-metallicity degeneracy \citep{worthey94}. 
 
 There are several ways to break this age-\Z degeneracy. While some studies employ pairs of colours in different wavebands \citep{li07, li13}, others use narrowband continuum photometry \citep{rakos07, rakos08, sreedhar12} or strength of different age-sensitive and {\it Z}-sensitive absorption line indices \citep{jorgensen97, jorgensen99, poggianti01}. One such set of absorption line indices was defined on the Lick/IDS system \citep[named after the Lick Image Dissector Scanner,][]{faber85}.

The H$\beta$ index is a widely adopted age-sensitive spectral index \citep{kuntschner98, kuntschner00, poggianti01, caldwell03}, while Iron ($\langle\rm{Fe}\rangle$\footnote{Throughout this paper we use $\langle\rm{Fe}\rangle$ to represent the average of the $Fe5270$ and $Fe5335$\AA\ lines.}) and Magnesium (Mg$_2$\footnote{The Mg$_2$ index is measured in the wavelength band 5154.125-5196.625 \AA.}) line indices are commonly used as a proxy for stellar \Z \citep{jorgensen97, jorgensen99, poggianti01, proctor04, sb06b}. A popular method to estimate stellar age and \Z involves plotting an age-sensitive index versus a {\it Z}-sensitive index together with a Single Stellar Population (SSP) model grid which translates the observed line strengths into mean age and \Z of the galaxy, respectively. Several such SSP models are available in the literature \citep{bc93, buzzoni95, weiss95, bressan96, vazdekis96, bc03, maraston05, vazdekis10}, and have been used to study the correlation of age and metallicity with other galaxy properties, such as stellar mass.
 
 The mass-metallicity relation extends across at least three orders of magnitude in stellar mass and a factor of 10 in gas-phase metallicity, though with a significant scatter \citep{tremonti04}. Various studies have explored the physical causes of the scatter in the mass-\Z relation, and the obvious culprit is found to be the observational uncertainty in estimating these quantities. Nevertheless, \citet{cooper08} used a sample of more than 57,000 star-forming galaxies from the Sloan Digital Sky Survey (SDSS), data release 4 to show that environment of galaxies can contribute $\sim 15\%$ of the observed scatter to the mass-\Z relation \citep[also see][]{mouhcine07}. The local density of galaxies dubbed ``environment" is known to have a strong impact on the evolution of galaxies, such that the optically-red, passively-evolving galaxies are found to preferentially reside in dense environments. Their disky, blue, star-forming counterparts on the other hand prefer intermediate or low density environments. This dichotomy is manifested in the well known morphology-density \citep{dressler80}, colour-density \citep[e.g.][]{blanton05} or star formation-density \citep[e.g.][]{balogh04} relations. 
 
 Galaxies in cluster cores are found to be more enriched relative to their counterparts in less dense environments \citep{skillman96,carter02,ellison09,peng14,wu17}. Using data from the SDSS it has been shown that at a given stellar mass the gas-phase metallicity of star-forming satellite galaxies is higher than star-forming central galaxies \citep{pasquali12,peng14}. Strangulation of low-mass galaxies, stripping of gas from the outer parts of the disk and external pressure of the intra-cluster medium (ICM) are the likely causes for these observations \citep{pasquali12}. 
 
 \citet{peng14}, however find that while the high-\Z satellites are found in denser environments relative to their low-\Z counterparts, no such trends are evident for star-forming central
 galaxies. Based on their findings, \citet{peng14} suggest that the observed trends are a consequence of enriched gas inflowing into satellite galaxies in dense environments. Analogous conclusions were drawn by \citet{darvish15} who studied a filamentary structure ($z\sim0.53$) in the COSMOS field. \citet{darvish15} find that galaxies in the filament are more metal-enriched ($\sim$ 0.1-0.15 dex) than a control sample of galaxies in voids.
 
 Enhanced rate of star formation in dense environments can also augment chemical enrichment in cluster galaxies at high redshift ($z\sim3$), but little correlation is observed between stellar \Z and environment at $z<0.5$ \citep{sheth06,panter08}. On the contrary, using a sample of $\sim 38,000$ galaxies from the SDSS data release 7, \citet{wu17} showed that the gas-phase \Z is weakly  dependent on environment below the turnover stellar mass ($M^*/M_{\odot} \sim 10^{10.5}$) upto which the metallicity is linearly correlated with the environment. Beyond the turnover mass, the  metallicity saturates, therefore becoming independent of environment. In light of the existence of mass-\Z relation, this observation of \citet{wu17} agrees well with the results of \citet{peng14}. 
  
 For galaxies in the Coma cluster ($z=0.023$), \citet{carter02} find a gradient in the Mg$_2$ and $\langle\rm{Fe}\rangle$ indices, which they attribute to pressure confinement of the  supernova ejecta by the ICM \citep[also][]{pasquali12}. Their suggestion is loosely supported by the ram-pressure stripping (RPS) model of \citet{gupta17}, who show that gas removal by RPS will primarily remove low-mass cluster galaxies from the star-forming population, leading to a cluster-scale metallicity gradient of -0.03 dex Mpc$^{-1}$ at $z=0.35$. \citet{gupta17}, therefore hypothesize that RPS is insufficient in explaining the observed \Z gradient in $z=0.35$ cluster, which would require ``self-enrichment due to gas strangulation''. In a study of galaxies in the nearby universe, \citet{hughes13} came to similar conclusions by studying the oxygen abundances and HI gas mass and gas fraction in 260 star-forming galaxies. Based on their analysis, \citet{hughes13} conclude that the mass-\Z relation is invariant to environment and originate only due to the internal evolutionary  processes in galaxies.
 
 Just like metallicity, the luminosity-weighted age of galaxies is also found to correlate with their environment. While older galaxies preferentially inhabit dense regions of space such as cluster cores, younger galaxies are abundantly found in the low-density environment of voids \citep{sb06a, annibali07, collobert06, trager08, pasquali10}. Several studies of clusters of galaxies have observed a trend of younger mean ages with increasing cluster-centric radius in the Coma \citep[$z=0.023$;][]{poggianti01, smith08, smith09} and Virgo \citep[$z=0.0036$;][]{michielsen08, toloba09} clusters. 
 
 Despite progress being made in understanding the impact of environment on the mass-\Z as well as age-\Z relations in the last few decades, a gap remains in the literature because of the availability of limited suitable samples. Most of the literature cited above either analyses the trends in over-dense cluster cores or in the almost empty void regions \citep[e.g.][]{jorgensen99,kuntschner00,moore01,poggianti01,caldwell03}. While the studies sampling the cluster core are biased towards passively-evolving galaxies populating dense environments, the ones using gas-phase metallicities often sample the star-forming galaxies only. But with the availability of wide-angle spectroscopic dataset from the SDSS it is now possible to cover these gaps and explore properties of galaxies covering a large range in stellar mass in a continuous range of environments.   
 
 In this work we employ data for the Coma supercluster spanning 480 square degrees \citep{SM10,SM18} on the sky to explore trends in luminosity-weighted age and stellar metallicity of galaxies with stellar mass and  environment. The Coma supercluster is one of the largest structures in the nearby Universe comprising two rich galaxy clusters: Coma (Abell~1656) and Abell~1367, connected by a network of large-scale filaments \citep{fontanelli84} and several small galaxy groups, making it an excellent site to investigate the age and \Z of its constituent galaxies.

 This paper is organised as follows. The observational data and procedure for selection of sample are described in the following section. In Sec.~\ref{s:age-z} we describe the methodology to estimate the luminosity-weighted age and stellar \Z of galaxies. In Sec.~\ref{analysis}, we explore the various factors affecting the age and metallicity of galaxies, discussing our observations in the context of existing literature in Sec.~\ref{discuss}, and finally concluding in Sec.~\ref{sum}. Throughout this paper we use $\Lambda$ cold dark matter concordance cosmology parameterised by $\Omega$$_{m}$=0.3, $\Omega$$_{\Lambda}$=0.7 and H$_{0}$=70 km s$^{-1}$ Mpc$^{-1}$ for determining distances and magnitudes.

\section{Data}
\label{data}

\begin{figure*} 
	\includegraphics[width=17cm]{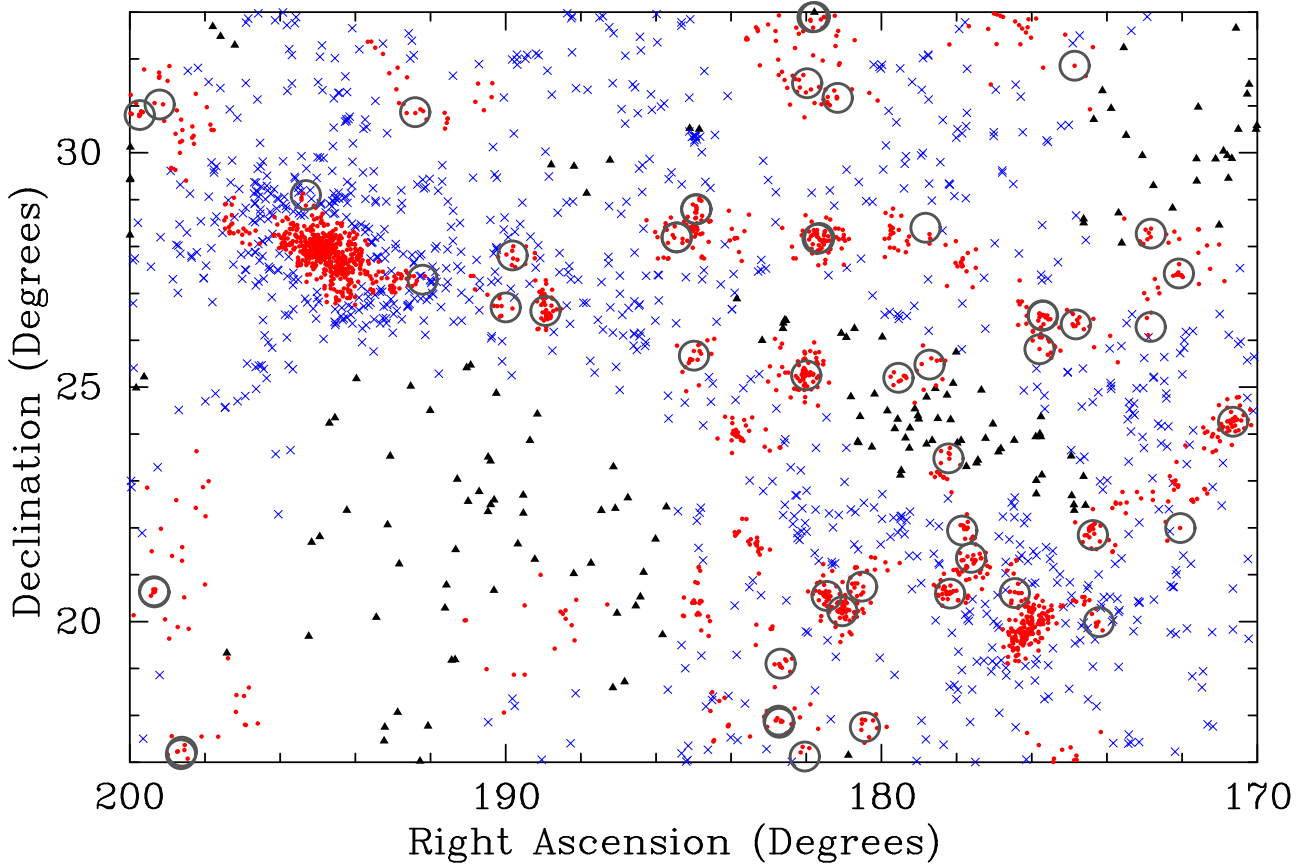}
	\centering
	\caption{A 2-d sky representation of Coma Supercluster. 2,934 galaxies out of 2,953 galaxies in the complete sample, for which environment information is available are shown. {\it Red points, blue crosses} and {\it black triangles} represent galaxies in clusters/groups, filaments and void regions, respectively. The {\it grey circles} represent galaxy groups identified in the NASA/IPAC Extragalactic Database (NED). }
	\label{scl}
\end{figure*}

 The data used in this paper are drawn from the data release 8 of the Sloan Digital Sky Survey Data \citep{york, aihara11}. The spectroscopic catalogue of the SDSS galaxies is complete to $r\leq17.77$ mag. We therefore implement this criteria together with the following to select galaxies for our sample:
 \begin{itemize}
  \item 170$^{\circ}$ $\leq$ RA $\leq$ 200$^{\circ}$,
  \item 17$^{\circ}$ $\leq$ Dec $\leq$ 33$^{\circ}$, and 
  \item radial velocity, $cz$ such that $4260\leq cz\leq9844$ km s$^{-1}$ ($0.0142\leq z \leq 0.0328$).
 \end{itemize} 
 The latter is chosen so as to include all galaxies within $\pm 3\sigma$ of the mean redshift of the Coma and the Abell~1367 clusters \citep[6973 and 6495 km s$^{-1}$, respectively;][]{rines03}. These data, covering $\sim 480$ square degrees on the sky are shown in Figure~\ref{scl}. For this sample, we adopt the stellar mass completeness limit of log $M^*/M_\odot = 9.05$ \citep[following fig.~8 of][]{weigel16}.
\begin{table}
	\centering
    \caption{Environment segregation of the samples}
	\label{tab:1}
	\begin{tabular}{lcc}
		\hline
        \hline
        Environment & Red sample & Complete sample\\
        \hline
        Cluster/group & 837 & 1810\\
        Filament & 364 & 963\\
	Voids & 57 & 161\\
 \hline
 \hline
	\end{tabular}
\end{table}
  \subsection{Properties of galaxies}
  The physical properties of galaxies as well as their metallicity ({\it Z}) and age in this work are drawn from the catalogues provided by the Max Planck Institute for Astrophysics (MPA) and Johns Hopkins University (JHU) group. The MPA-JHU catalogues make use of the spectra obtained using the $3^{\prime\prime}$ diameter fibre. Model spectra are fitted to the emission-line free regions of the observed spectrum. The residuals obtained after subtracting the fitted spectrum from the observed spectrum are then modelled as gaussians giving emission line measurements \citep{tremonti04}. In order to produce an absorption-line only spectrum, the fitted emission lines are subtracted from the observed spectrum. This work makes use of absorption line data only. The stellar masses in the MPA-JHU catalogue are calculated based on the methodology described in \citet{kauff03}, with the difference that the MPA-JHU pipeline makes use of the {\it ugriz} galaxy photometry instead of spectral indices. The \smass within the SDSS spectroscopic fibre aperture is estimated using fibre magnitudes whereas the total stellar mass uses model magnitudes. A Kroupa IMF is assumed. The output is provided as 2.5, 16, 50, 84 and 97.5 percentiles of the probability distribution function (PDF) of the logarithm of the stellar mass. In this work we make use of the median value of the stellar mass PDF. SDSS photometry catalogues list several different measures of magnitudes. Of these, model magnitudes, which are estimated by fitting a surface brightness profile to the two-dimensional image of a galaxy, usually provide the best available colour for extended objects. In this work we use the extinction corrected $g$ and $r$ band model magnitudes. 
 
 \subsection{Environment}
 \label{s:env}
Environment of galaxies can be broadly classified into two regimes: `local environment' ($\lesssim 1$ Mpc), which is often well quantified by measuring the local projected density using nearest neighbours,  and the `large-scale' environment which quantifies the $> 1$ Mpc density field around galaxies. Using a large sample of clusters, \citet{rines05} showed that beyond the virial radius, the local density at a given projected distance from the centre of the cluster can vary by as much as two orders of magnitude (their fig.~12). \citet{muldrew12} showed that environment metrics based on nearest neighbour density are well suited to measure the local environment in large halos, where as aperture-based methods are better representations of the large-scale environment of galaxies. Since we are interested in exploring the relation between galaxy properties and their large-scale environment in this paper, we adopt a method based on the latter methodology to quantify environment.
      
 In a recent paper, \citet{SM18} classified galaxies in the Coma supercluster into different environments. Here we use the same scheme to classify our galaxies into three distinct environments: 
 clusters/group, filament or voids\footnote{Since our data is sampled from DR8 unlike \citet{SM18}, environment information is missing for 19 galaxies of which 5 belong to the red sample.}. The number of galaxies inhabiting different environments for the giant and dwarf galaxies are presented in Table~\ref{tab:1}. For completeness, we summarise the definition of different environments below:
\begin{itemize}
\item {\it Cluster/group galaxies:} all galaxies identified to lie in groups or clusters. 
\item {\it Filament galaxies:} all galaxies that are within a radius of 1 Mpc of the filamentary spine.
\item {\it Void galaxies:} all other galaxies in our sample which are not selected in either of the above two categories.
 \end{itemize}
 The distribution of galaxies in different environments in the Coma Supercluster is shown in Fig.~\ref{scl}.
 \subsection{Age and \Z of galaxies}
 \label{s:age}
 The Lick system \citep{worthey94} is a set of twenty-one atomic and molecular absorption features which were initially measured in a sample of 460 nearby galactic stars in the 4000-6000 \AA\ waveband, and later updated with further observations by \citet{worthey97}.  
 All the stars were observed at Lick Observatory between 1972 and 1984 using the 3m Shane telescope and spectra obtained with the Cassegrain spectrograph and image dissector scanner (IDS) at moderate-resolution ( FWHM $\sim9$\AA). The first eleven Lick indices were defined using a sample of stars which were primarily K giants and subgiants in the solar-neighborhood \citep{faber85}. \citet{gorgas93} extended that sample to include dwarf stars (F and G type) in the field and clusters
 to measure the eleven indices. \citet[][W94 henceforth]{worthey94} further extended the stellar sample to include M type dwarfs and giants, and several hot stars to span the old, metal-rich populations along with the young stars 0.5-2 Gyr old. W94 also supplemented the original list of eleven absorption line indices with ten new features, thus constituting the whole Lick/IDS system.

 In this work we adopt three line indices,  H$\beta$ (4847.875 \AA- 4876.625 \AA) as an indicator of age, and mean of  Fe5270 (5245.650 \AA- 5285.650 \AA) and Fe5335 (5312.125 \AA- 5352.125 \AA) indices, $\langle Fe \rangle$ as an indicator for \Z of galaxies in our sample. We note that all absorption line indices are measured from galaxy spectra after subtracting all emission-lines with signal-to-noise ratio (SNR) $\geq 3\sigma$. The reader is directed to \citet{brinchmann04} and \citet{tremonti04} for a detailed description of the methodology adopted for the measurement of spectral indices from the SDSS spectra.  
 
 Although these indices have been used in the literature to analyse populations of galaxies, especially in the passively-evolving galaxies in galaxy clusters, our work is distinguishable because we have analysed `all' types of galaxies i.e. emission-line and absorption-line systems, spanning a wide range in stellar mass and environmental density homogeneously. This uniform treatment of data allows us to get an insight into the evolution of galaxies not just in dense clusters, but in the much larger supercluster. In order to meet
 our science goals it is therefore essential to include all types of galaxies even if some of them may have high uncertainties. However, in order to test for biases in our results arising due to high observational uncertainties, we perform the analysis by segregating our data into two sub-samples: 
\begin{itemize}
 \item {\bf Complete sample} of 2,953 galaxies, and
 \item  {\bf Red sample}: A reduced subset of the complete sample comprising 1,263 galaxies with {\it $< 20$ per cent uncertainty in all the three absorption line indices} used in this work.
\end{itemize}
 
\begin{figure}
	\includegraphics[width=8.5cm]{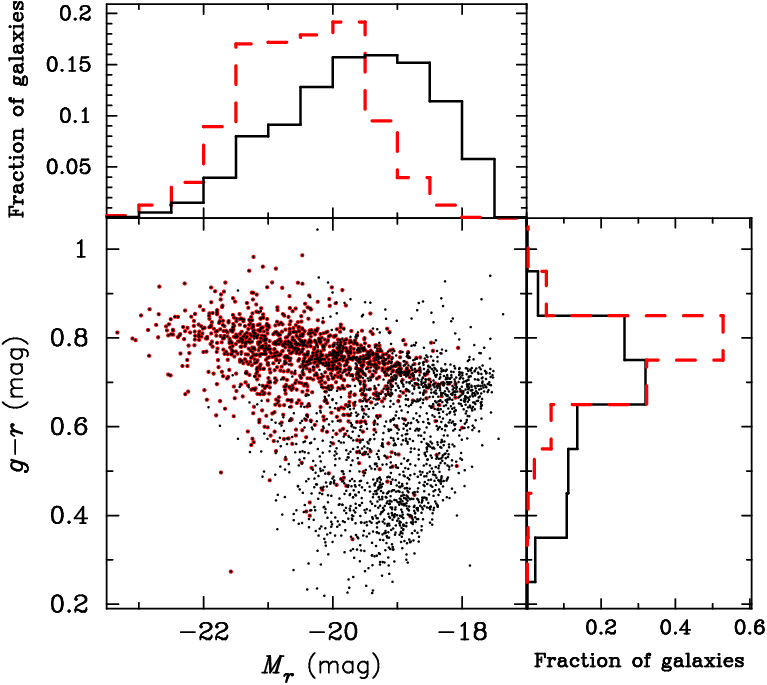}
	\centering
	\caption{ This figure shows the distribution of all galaxies in the colour-magnitude plane, and the respective distributions of the $(g-r)$ colour and $M_r$ magnitude for the complete {\it (black points, solid lines)} and red {\it (red points, dashed lines)} sub-samples, respectively. }
	\label{cmd}
 \end{figure}

 The distribution of all the galaxies used for analysis in this paper in the colour-magnitude plane is shown in Fig.~\ref{cmd}. As expected, most of the galaxies on the red sequence are strong absorption line systems, and therefore belong to the lower-uncertainty red sub-sample.  

\begin{figure*}
	\includegraphics[width=17cm]{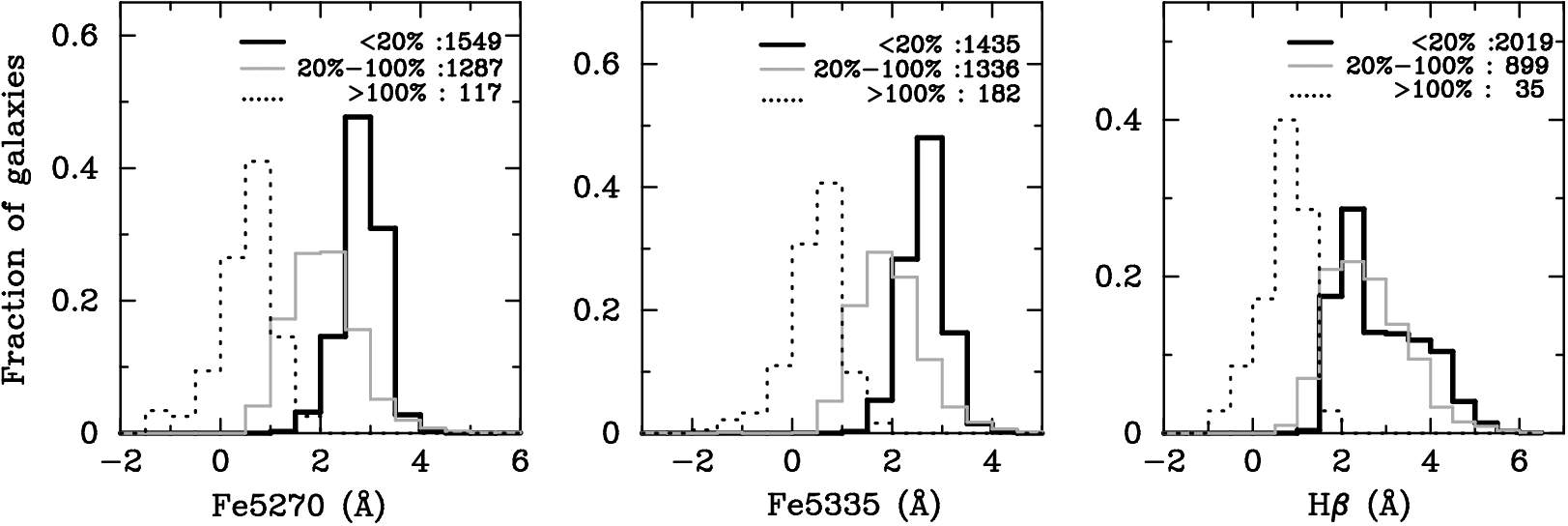}
	\centering
	\caption{This figure shows the distributions of absorption line indices for the complete sample of 2,953 galaxies segregated according to the relative measurement errors as indicated in the legend. The number of galaxies contributing to each distribution is also mentioned.  In general, larger absorption line equivalent width is synonymous with lower measurement error.}
	\label{fig:error_indices}
 \end{figure*}

 In Fig.~\ref{fig:error_indices} we show that the smaller `red' sub-sample is dominated by massive red galaxies, which are passively-evolving systems with low uncertainty in absorption line measurements. A corollary of this observation is that the star-forming galaxies with spectra dominated by emission lines are found to have the highest uncertainty in the absorption line indices required to estimate their age and metallicity. Specifically, the typical uncertainty in the H$\beta$ and $\langle Fe \rangle$ indices are 0.4 \AA\ (0.25 \AA) and 0.3 \AA\ (0.2 \AA) in the complete (red) sub-samples, respectively. Furthermore, $90$ per cent of the galaxies in the complete (red) sub-sample have $<29$ per cent ($<16$ per cent) and $<34$ per cent ($<11$ per cent) uncertainty in the H$\beta$ and $\langle Fe \rangle$ indices, respectively. 

\section{Estimation of age and metallicity}

\label{s:age-z}

 In this work we use Single Stellar Population (SSP) models, which transform the observed line indices into luminosity-weighted mean age and mean metal content of galaxies. In particular, we use the SSP models described in \citet[][hereafter V10]{vazdekis10}. V10 make use of the solar-scaled theoretical isochrones of \citet[][hereafter Padova isochrones]{girardi2000} and cover a wide range of parameters and initial mass functions (IMFs). The Padova isochrones are based on theoretical stellar evolutionary tracks for many low and intermediate mass stars ($0.15-7 M_\odot$) covering a wide range in metallicity ($0.0004\leq Z\leq0.03$). The stellar tracks are presented at very small mass intervals which allows for a detailed mapping of the Hertzsprung-Russell Diagram, which results in a detailed derivation of isochrones \citep{girardi2000}. 
  
  V10 compute the SSP spectral energy distributions (SED) by integrating the spectra of stars along the isochrone taking into account their number per mass bin in accordance with the adopted IMF. They make use of stars in the Medium-resolution Isaac Newton Telescope Library of Empirical Spectra (MILES) \\database, which have a good stellar atmospheric parameter coverage, and compute their predictions for different IMFs.
   
 Here we use the predicted line strengths based on the work of V10 from the MILES database. We have employed fourteen constant age 
 (0.79, 0.89, 1.00, 1.12, 1.78,  2.82, 3.16, 4.47, 5.01, 5.62, 6.31, 7.08, 7.94 and 14.12 Gyr) and six constant metallicity isochrones (-1.71, -1.31, -0.71, -0.4, 0.0, 0.22 [M/H]). The bimodal IMF (slope $=1.3$) closely resembles the Kroupa IMF \citep{kroupa01}, and the SSP SEDs obtained for these IMFs do not show significant differences (V10). Furthermore, the effects of varying the slope of the IMF on the SSP SEDs are far less significant for the bimodal IMF relative to those arising from varying the slope of the unimodal IMF (V10). We therefore adopt the bimodal IMF \citep{vazdekis96} in this work.
 
 The age and \Z of galaxies are estimated by interpolating from the nearest model points. However, for 98 (503) galaxies in the red (complete) sub-sample which lie outside the model grid, the age and metallicity of galaxies are estimated by extrapolating the constant age and \Z isochrones. In regions where the isochrones cannot be extended, i.e. $< 0.7943$ Gyr and $> 14.1254$ Gyr in age, and $Z < -1.71$ and $Z > +0.22$ for metallicity, galaxies are assigned an age of 0.05 Gyr and 20 Gyr, and a metallicity value of $\pm$4, respectively. In order to draw statistically valid results, these galaxies are excluded from further analysis.
 
 
 In order to compute the uncertainty in the age and \Z due to the measurement error in the H$\beta$ and $\langle Fe \rangle$ indices, the index values are perturbed with their uncertainties and age and \Z are recomputed. This gives us the upper and lower bounds in both the quantities for each galaxy. We note that several perturbed index values fall outside the model grid, either due to the measured value being very close to the grid boundary or having large uncertainty associated with it. In such cases, age and \Z bounds are estimated via extrapolation by extending the model isochrones where possible. The galaxies for which the isochrones can not be extended, are excluded from further error analysis. 
 
 Following \citet{jorgensen99} and \citet{poggianti01}, the uncertainties in age and \Z are then calculated by taking half of the difference between the upper and lower limits for each galaxy. In the complete (red) sub-sample, age and \Z bounds are well estimated for $62$ per cent ($74$ per cent) and $69$ per cent ($85$ per cent) of the galaxies, respectively. The fractions are naturally higher for the red sub-sample, owing to smaller uncertainty in the indices and the sub-sample consisting mainly of those galaxies which lie well within the model grid. Based on these galaxy fractions, the median absolute uncertainty in age is found to be 1.7 Gyr (2.4 Gyr), and in \Z is 0.26 dex (0.2 dex) for the complete (red) sub-sample, respectively. These uncertainties are comparable to those found by \citet{poggianti01} for the central region of the Coma cluster using data from the William Herschel Telescope. We present the age and \Z, and the respective uncertainties obtained for them for all the galaxies in our sample in Table~\ref{data-table}, which is available online in its entirety. The columns are: (i) Right Ascension (J2000), (ii) Declination (J2000), (iii) redshift, (iv) age (Gyr), (v) and (vi) upper and lower limits of age (Gyr), (vii) \Z (dex), (viii) and (ix) upper and lower limits of \Z (dex), respectively. 
 
 We note however, that the uncertainty in age varies with the age of the population, due to the logarithmic nature of the grid used to compute the ages and \Z \citep[Fig.~\ref{fe-hb}; also see fig.~8 of][]{poggianti01}. The typical uncertainty in age is found to be 1 Gyr (0.7 Gyr) for galaxies $< 3$ Gyr old for the complete (red) sub-sample. For older populations, the uncertainty is $\sim 3.4$ Gyr for both the sub-samples. The red sub-sample, chosen to have lesser uncertainty in the indices, is largely dominated by older systems with age $> 3$ Gyr (Fig.~\ref{fe-hb}). On the other hand, $\sim 65$ per cent of the galaxies in the complete sub-sample are $< 3$ Gyr old, consequently giving slightly higher age uncertainty of 2.4 Gyr for the red sub-sample when all galaxy populations are taken into account.

\begin{table*}\centering
\caption{The position, age and \Z for all the galaxies in the sample analysed here.}
\label{data-table}
\begin{tabular}{ccccccccc}
\hline
Right Ascension & Declination & Redshift & Age & Age Upper limit & Age lower limit & \Z & \Z upper limit & \Z lower limit \\
(J2000) & (J2000) & & (Gyr) & (Gyr) & (Gyr) & (dex) & (dex) & (dex) \\ 
 \hline\hline
178.312 & 20.736 & 0.0252 & 2.2 & 3.1 & 1.4 & -0.12 & - & -0.39 \\ 
191.190 & 27.500 & 0.0314 & 1.8 & 2.6 & 1.2 & -0.25 & - & -0.55 \\ 
197.003 & 28.081 & 0.0192 & 2.7  & 5.1 & 1.9 & -0.36 & -0.07 & -0.64 \\ 
189.744 & 27.564 & 0.0224 & 2.1 & 2.9 & 1.4 & -0.16 & - & -0.45 \\ 
179.050 & 21.394 & 0.0255 & 2.1  & 2.9 & 1.5 & -0.34 & 0.04 & -0.61 \\ 
180.642 & 20.815 & 0.0223 & 1.7 & 2.7 & 1.1 & -0.07 & - & -0.44 \\ 
195.960 & 28.054 & 0.0210 & 2.4 & 4.1 & 1.5 & -0.12 & - & -0.42 \\ 
181.856 & 28.530 & 0.0303 & 1.9 & 2.9 & 1.2 & 0.04 & - & -0.31 \\ 
181.500 & 27.639 & 0.0285 & 2.4  & 4.1 & 1.6 & -0.31 & 0.08 & -0.61 \\ 
177.888 & 27.669 & 0.0290 & 1.8 & 3.1 & 1.1 & 0.20 & - & -0.23 \\ 
182.909 & 19.845 & 0.0223 & 2.5  & 4.0 & 1.8 & -1.12 & -0.85 & -1.34 \\ 
\hline
\end{tabular}
\end{table*}

 

\section{Analysis and results}
\label{analysis}

 It is well known that massive, passively-evolving galaxies reside in dense environments. Therefore, in order to analyse the trends in age and metallicity with environment and stellar mass, it is essential to decouple the latter two properties of galaxies. In this work we accomplish this by splitting each sub-sample into bins of \smass 
 and environment, assuming insignificant evolution of galaxies within each sub-sample.
 
 We split our data into two bins of \smassa, viz. `dwarfs' are defined as galaxies having ${\rm log}~M^*/M_\odot < 10$ and $> 10$ for the `giant' galaxies, respectively. While 24 per cent of galaxies in the red sample are classified as dwarfs, almost 62 per cent of galaxies in the complete sample appear in this class. Fig.~\ref{cmd} shows the distributions of the \gr colour for the complete and the red sub-sample, respectively. As expected, galaxies in both the sub-samples become redder with increasing $r$-band luminosity, which is a good proxy for $M^*$. In particular, the lack of low-mass, blue galaxies, especially in the red sub-sample is evident, once again justifying the need to analyse the complete sample including galaxies with higher measurement uncertainties in the spectral indices.
  
\begin{figure*} 
   \includegraphics[width=18cm]{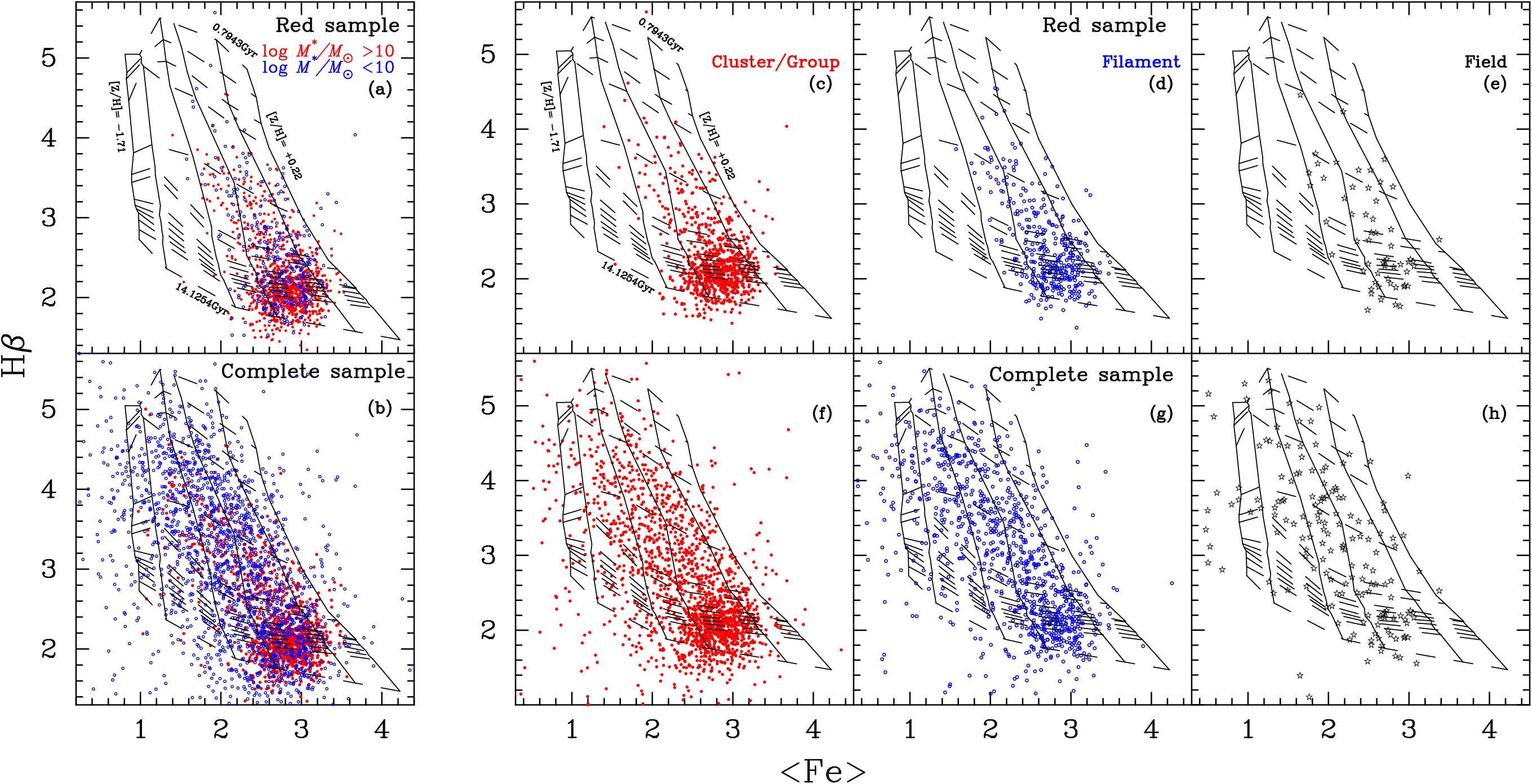}
    \caption{Distribution of galaxies in the H$\beta$-\fe~plane for the red {\it (top)} and complete {\it (bottom)} sub-samples. The {\it blue circles} and {\it red points} in panels (a) and (b) represent the 
    dwarf and giant galaxies in the two sub-samples, respectively. Overplotted are constant age {\it (dashed)} and constant metallicity {\it (solid)} isochrones. The age and metallicity values corresponding to 
    the extreme contour levels are marked. Panels (c) to (h) represent the distribution of galaxies in clusters, filaments and voids in the red {\it (panels c-e)} and complete {\it (panels f-h)}, respectively.    }
     \label{fe-hb}
\end{figure*}

 In Fig.~\ref{fe-hb} we show the distribution of galaxies for the red and complete sub-samples in the H$\beta$-\fe~plane, along with the constant age and \Z isochrones mentioned in Sec.~\ref{s:age-z}. Most of the passively-evolving giant systems have large absorption in \fe~and lower absorption in \hb, indicating high metallicity and lower star formation in the recent past, respectively. The dwarfs in the red sample follow the giants. However, the dwarfs in the complete sample span the entire \hb-\fe~plane, suggesting a wide range of star formation histories. 
 
 Segregation with environment (panels (c)-(h) in Fig.~\ref{fe-hb}) shows that most of the cluster galaxies are concentrated in the bottom-right quadrant dominated by massive, passively-evolving systems with a small but non-negligible fraction of the remaining galaxies spanning the rest of the \hb-\fe~plane. In filaments and voids however, galaxies span the entire H$\beta$-\fe~plane with some traces of an excess in the bottom-right quadrant in the filaments. The consequence of such a distribution in the context of the \smass distribution and environment of galaxies is discussed in the following sections.

\subsection{Disentangling the effects of age and $M^*$ on \Z}

  \begin{figure} 
  \includegraphics[width=8cm]{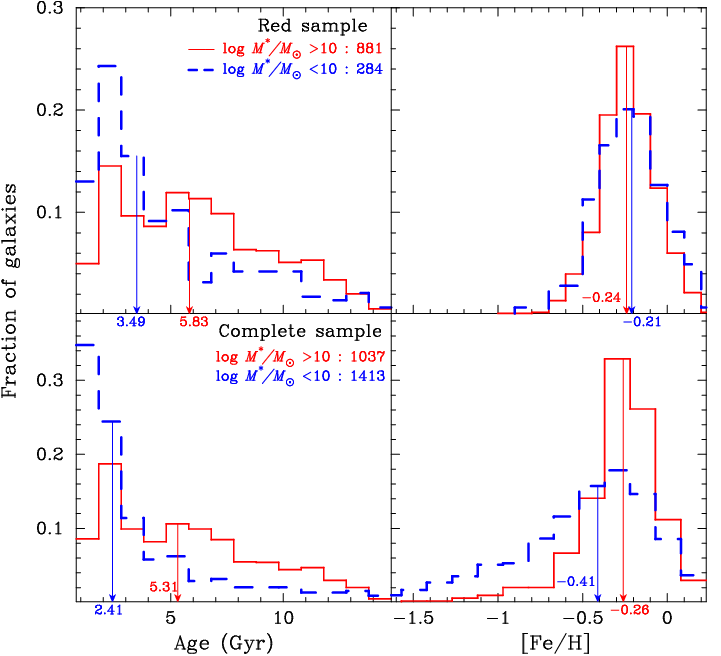}
   \caption{ The age {\it (left)} and metallicity {\it (right)} distributions of galaxies in the red {\it (top)} and complete {\it (bottom)} sub-samples, respectively. Each sub-sample is further split into dwarfs and giants. The bottom-pointing arrows in each panel represent the median age and metallicity of the distributions. }
     \label{age-z-dist}
 \end{figure}

 \begin{figure*}
 \includegraphics[width=18cm]{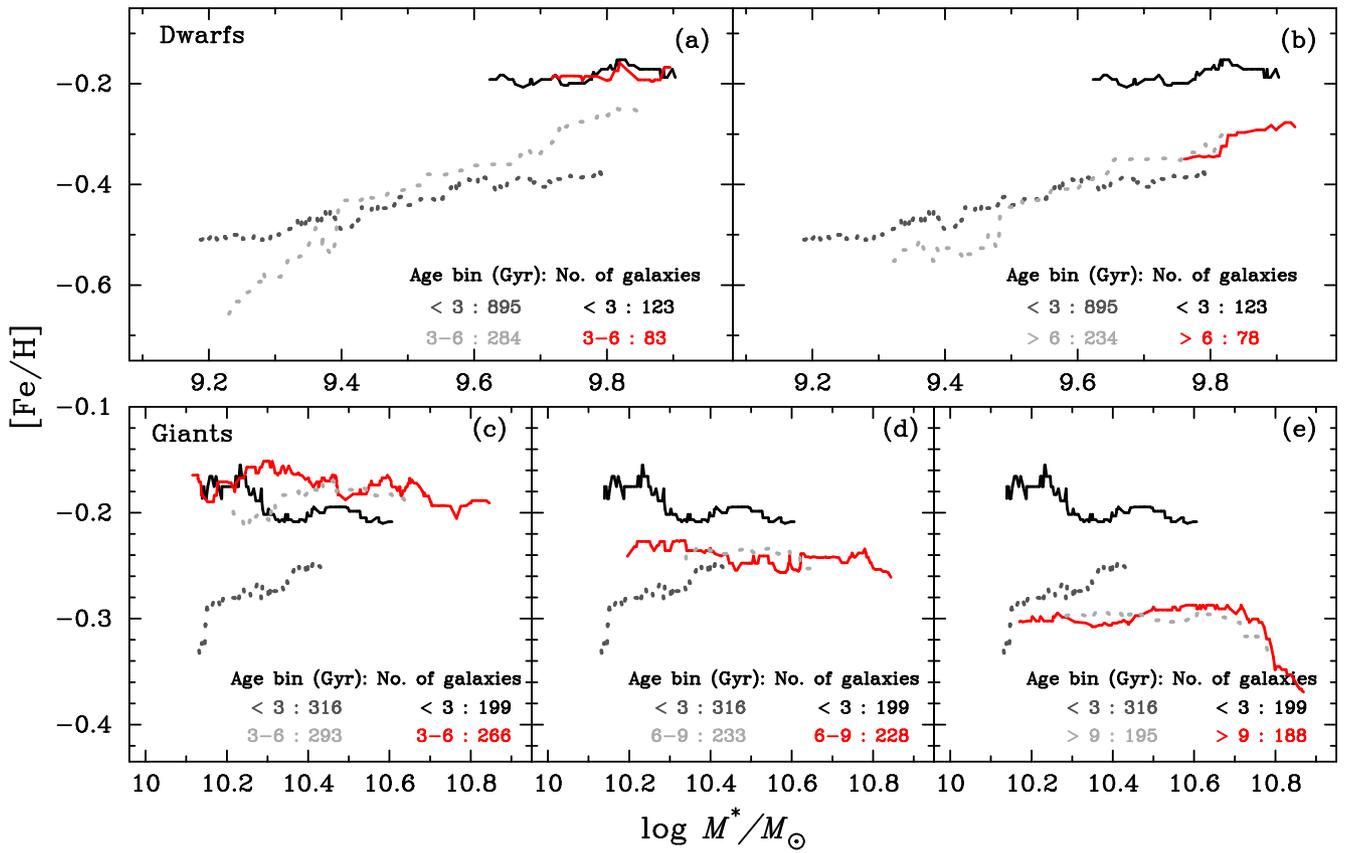}
     \caption{The median trend in the metallicity of {\it (top)} dwarf and {\it (bottom)} giant galaxies as a function of \smassa. The solid lines represent the red sub-sample, while the complete sample is shown using the dotted lines. The {\it black lines (solid for red sub-sample and dotted for complete sample)} in each panel represent the metallicity distribution of the youngest galaxies in the specified mass bin, which is compared to the sub-sample of older galaxies {\it (solid red for red sub-sample and dotted grey line for complete sample)}. The age of the sample to be compared with the youngest galaxies increases from left to right as indicated by the age range mentioned in each panel along with the number of galaxies contributing to each curve.}
    \label{z-smass}
  \end{figure*}
  
  In this section we study the metallicity of galaxies as a function of their stellar mass and luminosity-weighted age. In Fig.~\ref{age-z-dist} we show the distributions of age and \Z for the dwarf and giant galaxies along with the median for each distribution. Both samples show statistically significant differences in the distribution of the age of the dwarf and giant galaxies, but negligible differences in the distribution of their {\it Z}. These inferences are drawn on the basis of the outcome of the Kolmogorov-Smirnov (KS) statistical test, which tests for the likelihood that the two samples being compared originate from the same parent distribution. The KS test probabilities in favour of the hypothesis for the red and the complete samples for age and \Z of galaxies in different mass bins and environments are shown in Tables~\ref{a1} and \ref{a2}, respectively.
  
 \begin{table}\centering
    \caption{KS test probabilities for age and metallicity distributions of galaxies in different mass range and environments in the red sub-sample.}
	\label{a1}
	\begin{tabular}{ccc} 
        \hline
        \hline
       \smass & Age & [Fe/H]\\
       \hline
       	Dwarfs \& Giants & 2.3E-12 & 8.0E-02\\
        \hline
        \hline
       Environment & & \\
        \hline
       	Cluster/Group & &\\
        \& Filament & 5.9E-05 & 8.8E-01\\
        \hline
		Cluster/Group & &\\
        \& voids &3.0E-03 & 8.0E-02\\
        \hline
        Filaments \& voids & 2.5E-01 & 1.9E-01\\
        \hline
       \hline
	\end{tabular}
\end{table}

\begin{table}\centering
    \caption{KS test probabilities for age and metallicity distributions of galaxies in different mass range and environments in the complete sample.}
	\label{a2}
	\begin{tabular}{ccc} 
        \hline
        \hline
       \smass & Age & [Fe/H]\\
       \hline
       	Dwarfs $\&$ Giants & 4.8E-62 & 5.8E-47\\
        \hline
        \hline
       Environment & & \\
        \hline
       	Cluster/Group & &\\
        $\&$ Filament & 3.2E-14 & 7.7E-05\\
        \hline
		Cluster/Group & &\\
        $\&$ voids &6.3E-07 & 1.3E-05\\
        \hline
        Filaments $\&$ voids & 3.0E-01 & 2.0E-02\\
        \hline
       \hline
	\end{tabular}
\end{table}

  Taking inspiration from \citet{poggianti01}, we split our sample into four bins in age: < 3 Gyr (very young galaxies), 3-6 Gyr (intermediate), 6-9 Gyr (intermediate old) and >9 Gyr (very old). In Fig.~\ref{z-smass} we show the median trend in metallicity as a function of stellar mass for the dwarf and giant galaxies in each bin of age for the red and the complete sub-samples, respectively. For the dwarf galaxies we have combined the intermediate-old and very old galaxy populations due to low number statistics. Both the dwarf and giant galaxies show a continuous decline in metallicity with increasing age. Although counter-intuitive, this result is explainable in a scenario where galaxies grow by accreting smaller satellites and cold gas from the large-scale structure in which they are embedded. Such cold-mode accretion will feed low metallicity gas, thereby reducing the mean metallicity of the progenitor galaxy \citep[e.g.][]{dsouza18}. 
 
 Galaxies can lose enriched inter-stellar medium (ISM) via galactic outflows, for instance, in supernovae explosion, resulting in a decline of the mean metallicity of galaxies \citep{chisholm18}.
 This hypothesis is clearly observed for the red sub-sample, where the \Z of galaxies is found to decline with increasing age, irrespective of their \smassa. This trend is duplicated for the complete sample of giant galaxies, and albeit with some scatter, also for the complete sample of dwarf galaxies. At fixed age however, \Z is mostly independent of $M^*$ for the dwarf and giant galaxies alike, except for the most-massive, oldest galaxies which show a sharp decline in their {\it Z}, likely caused by excessive cold-mode accretion in the central dominant galaxies in clusters and groups.  
   
 \subsection{Age and \Z of dwarfs and giants in different environment}
 \label{s:z-env}
 
  \begin{figure*}
 \includegraphics[width=18cm]{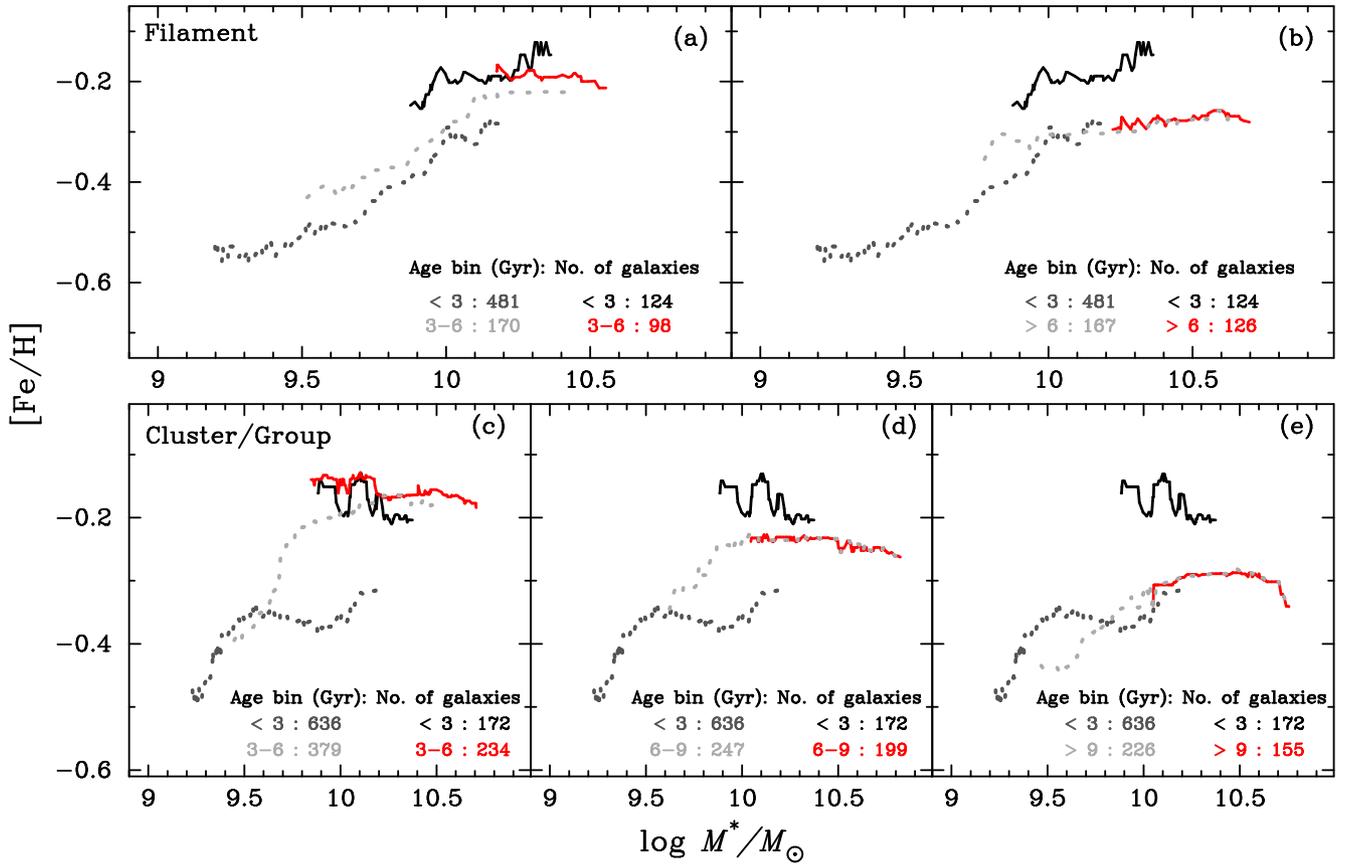}
    \caption{The median trend in the metallicity for galaxies in {\it (top)} filaments and {\it (bottom)} clusters or groups. The line styles are same as in Fig.~\ref{z-smass}. Void galaxies are not shown due to low number statistics. }
    \label{z-smass-env}
\end{figure*}
 
  \begin{figure} 
  \includegraphics[width=8cm]{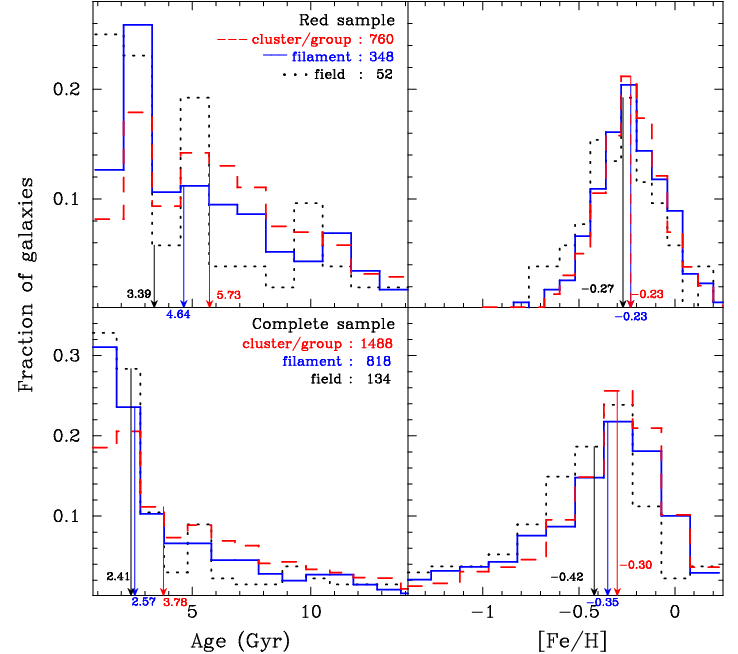}
  \caption{Same as Fig.~\ref{age-z-dist}, but for each sample divided on the basis of environment as classified in \citet{SM18}. Galaxies become older and metal-rich as the density of environment increases from field to filaments to clusters or groups. }
  \label{env-seg}
\end{figure}

\begin{table}
	\centering
    \caption{The fraction of young, old and metal-rich and metal-poor galaxies among each category of galaxies in the two samples.}
	\label{frac}
	\begin{tabular}{lcc}
        \hline
         &  Red sub-sample   &\\       
       \hline
        & $\%$Old   & $\%$Metal rich \\
        &  (>5Gyr) &  ([Fe/H]>-0.3)\\
        \hline
        Dwarfs &  36.3 &  65.8\\
	Giants &  59.8 & 66.6\\
    \hline
        Galaxy Cluster/Group &  58.3 &  67.6\\
	Filament &  47.1 & 65.2\\
        Field &  40.4 &  55.8\\
 	\end{tabular}
	
	\centering
	\begin{tabular}{lcc}
		\hline
        \hline
         &  Complete sample   &\\       
       \hline
        &  $\%$Old &  $\%$Metal rich \\
        &  (>5Gyr) &  ([Fe/H]>-0.3)\\
        \hline
        Dwarfs  & 22.6 &  36.4\\
	Giants  & 52.6 &  59.8\\
    \hline
        Galaxy Cluster/Group  & 40.7  & 49.5\\
	Filament  & 27.5  & 42.7\\
        Field  & 23.9  &32.8\\
 \hline
	\end{tabular}
\end{table}

 In Fig.~\ref{z-smass-env} we show the median trend in metallicity as a function of stellar mass for galaxies in clusters and filaments in each bin of age for the red and the complete sub-samples, respectively. The void galaxies could not be shown here due to low number statistics. Just like Fig.~\ref{z-smass}, the intermediate-old and very old populations are combined together for the filament galaxies due to low number of galaxies in these bins. Fig.~\ref{env-seg} shows the distribution of age and \Z of galaxies in different environments for the two sub-samples. The previously observed trend of decreasing \Z with increasing age is seen in all sub-samples, except the complete sub-sample of filament galaxies. Specifically, while 56 per cent of galaxies in voids are metal-rich ([Fe/H]$>-0.3$), the fraction increases to 65 per cent for filaments, and further to 68 per cent for the cluster galaxies in the red sub-sample, suggesting that dense environments rich in passively-evolving galaxies also have more metal-rich galaxies. However, we must also consider the fact that dense environments comprise of more massive galaxies. In order to test the effect of environment on \Z of galaxies independent of stellar mass, we create sub-samples of galaxies in two stellar mass ranges for different environments. For massive galaxies ($M^*/M_{\odot}> 10^{10.5}$) the median \Z is found to be -0.26 (-0.26), -0.25 (-0.24) and -0.28 (-0.28) in clusters, filaments and field for the complete (red) sub-samples, respectively. Similarly, for the intermediate mass galaxies ($10^{10} \leq M^*/M_{\odot} < 10^{10.5}$) the same is -0.24 (-0.22), -0.26 (-0.22) and -0.36 (-0.27), respectively. This analysis implies that the \Z of void galaxies is statistically different from the \Z of galaxies in clusters or filaments.
 
 We also performed a similar analysis for the age of galaxies in    different environments in the two mass-matched sub-samples. The median age for galaxies in clusters, filaments and voids is found to be 6.79 (6.91), 6.01 (6.17) and 5.58 (5.28) Gyr for the massive galaxies, and 4.74 (5.45), 3.31 (4.55) and 2.76 (2.55) Gyr for the intermediate mass galaxies in the complete (red) sub-samples, respectively. Based on these data, statistically there is no age difference in the massive galaxies in clusters or filaments, but the void galaxies are younger by $\sim 0.5$ Gyr. The less massive galaxies on the other hand are clearly different in their age, such that the cluster galaxies are older than the filament galaxies by 1-1.5 Gyr, which in turn are older than their counterparts in the voids by $\lesssim 1$ Gyr. In general, we find that about two-thirds of all giant and dwarf galaxies have [Fe/H]$>-0.3$, but even though 64 per cent of the dwarfs are younger than 5 Gyr, the same applies to 40 per cent of the giant galaxies only (Table~\ref{frac}). 

 Together with these analyses and Figs.~\ref{z-smass} and \ref{z-smass-env}, we conclude that the \Z of galaxies in the Coma supercluster is primarily dictated by their age. \smass and environment have a second order effect on the \Z such that younger, lower mass galaxies also have lower {\it Z} as found by other studies as well \citep[e.g.][]{poggianti01}. 


\section{Discussion}
\label{discuss}
 
 We use the Lick indices' measurements for galaxies to study the correlation between age, stellar {\it Z}, stellar mass and environment of galaxies in the Coma supercluster. While the giant galaxies are mostly old and metal-rich, their dwarf counterparts are relatively younger and metal-poor (Table~\ref{frac}, Fig.~\ref{age-z-dist} and \ref{m-z}). These observations are a manifestation of the downsizing scenario observed elsewhere \citep{heavens04, gallazzi05, jimenez05, thomas05, neistein06, fernandes07, panter07, fontanot09, 
 thomas10}, viz. galaxies at fixed \smass are found to be forming fewer stars with decreasing redshift. In the following we analyse our results in the light of information from the existing literature, also describing the shortcomings of the analyses presented in this paper.  
 
 \subsection{Age-\Z anticorrelation}
 In agreement with the literature \citep[e.g.][]{jorgensen99, poggianti01, kuntschner00, price11}, we find that the (luminosity-weighted) age and stellar \Z of galaxies are anti-correlated, such that the younger galaxies tend to be more metal-rich than their older counterparts. These trends are observed in all environments (Fig.~\ref{z-smass-env}), and for the dwarf ($M^*<10^{10} M_{\odot}$) and giant galaxies ($M^*>10^{10} M_{\odot}$; Fig.~\ref{age-z-dist}) alike. This trend was previously observed for the passively-evolving elliptical and S0s in clusters. In our work presented here, we observe the same even after incorporating the late-type galaxies and sampling a continuous range of environments present in the Coma supercluster. 
  
 Such an anticorrelation where errors in both age and \Z are correlated \citep{kuntschner2001}, is difficult to assess because an underestimation in one leads to an overestimation of the other \citep{trager2000}. Hence this trend may conspire to give a tight correlation between age or \Z sensitive indices and other global properties such as the stellar mass \citep[Fig.~\ref{age-z-dist};][]{tremonti04,gallazzi05} or luminosity of galaxies \citep{bell00,poggianti01}. Although it is worth noting that if correlated errors were the only cause, such a relation would have been more apparent in low S/N galaxies, i.e. the ones with the highest errors. But the fact that the anticorrelation is unambiguously observed for the red sub-sample (i.e. high S/N galaxies), at all masses and environments implies that for our sample the age-\Z anticorrelation is not an artefact, in agreement with \citet{poggianti01} who employed data for two physically separated regions in the Coma cluster.
 
 \subsection{The mass-\Z relation}
 
  \begin{figure} 
 \includegraphics[width=8cm]{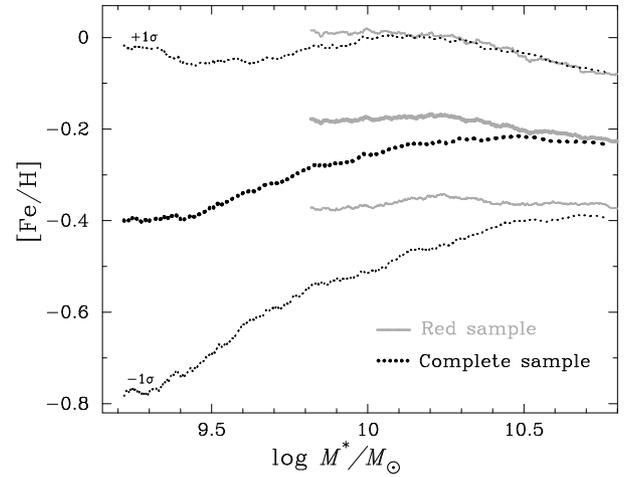}
 \caption{[The median Fe/H] ratio as a function of \smass for the complete {\it (black)} and red {\it (grey)} sub-samples, respectively. The lines representing the $\pm1\sigma$ limit
 are shown on either side of the median relation. Notice that the high S/N (red) sub-sample comprises mainly of massive galaxies. }
 \label{m-z}
 \end{figure}
 
 In Fig.~\ref{m-z} we show the \Z of galaxies as a function of their \smass for the complete and red sub-samples, along with 1-$\sigma$ deviations in the relation of running median. 
 The observed trend is similar to that observed elsewhere \citep[e.g.][]{bell00,tremonti04,gallazzi05}, i.e. the metallicity increases with stellar mass or luminosity, and then saturates for the massive galaxies ($M^*/M_{\odot} \gtrsim 10^{10.5}$).
 This relation reflects the importance of stellar mass in determining the composition of galaxies.
 Also, the scatter in the $M^*$-\Z relation decreases with increasing mass.
 The large scatter in the lower mass galaxies is partly due to the larger uncertainties in the age and \Z of low-mass galaxies. However, even at fixed stellar mass the scatter is larger than the uncertainties in {\it Z} \citep{gallazzi05}.

 In a sample which spans a range of environments, such scatter is expected due to the variety of galaxies contributing to the relation \citep{cooper08}. For instance, despite having very different metallicities, dwarf elliptical galaxies and irregulars will both contribute to the $M^*$-\Z relation at log $M^*/M_{\odot} \sim 9$, thereby adding to the scatter caused by the variations in the intrinsic properties of galaxies (at fixed mass).
 
 The mass-\Z relation also has a strong dependence on the star formation rate (SFR) of galaxies. These quantities form the ``fundamental metallicity relation (FMR)" \citep{lara10, mannucci10}. Furthermore, using data from the ALFALFA atomic gas survey, \citet{bothwell13} showed that galaxies in the nearby Universe obey an HI FMR, between stellar mass, gas-phase metallicity and the HI mass. \citet{bothwell13} hypothesize that the HI FMR relation is more fundamental and drives the relation between SFR, \smass and metallicity because the dependence of \Z on HI mass does not saturate like it does for the SFR. \citet{bothwell13} also find that at fixed stellar mass, galaxies with higher atomic gas mass have relatively lower gas-phase {\it Z}.   

  \subsection{Metallicity of galaxies in different environments}
  
 In Sec.~\ref{s:z-env} we observed that environment and stellar mass play a secondary role in determining the \Z of galaxies. The \Z of galaxies is primarily determined by their age, such that the older galaxies have lower \Z than their younger counterparts. We now explore the impact of environment on the metallicity of galaxies. 
   
 \begin{figure} 
 \includegraphics[width=8cm]{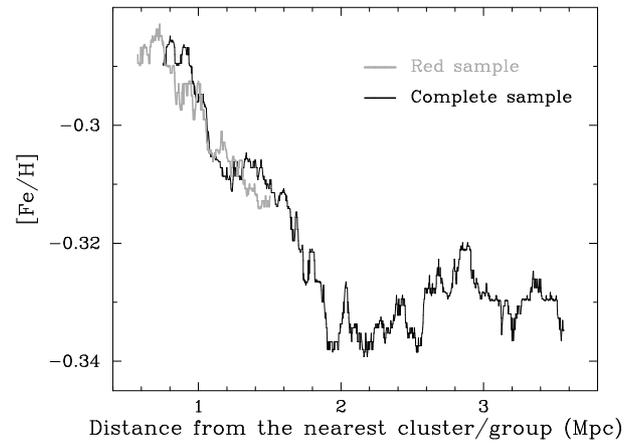}
 \caption{This figure shows the median [Fe/H] ratio for cluster galaxies as a function of their distance from the centre of the nearest cluster centre for the complete {\it (black)} 
 and red {\it (grey)} sub-samples, respectively. In the complete sample, \Z decreases with increasing cluster-centric distance saturating to [Fe/H]$\sim -0.33$ at $\gtrsim 2$ Mpc from 
 the cluster centre.}
 \label{z-env}
 \end{figure}

 In Fig.~\ref{z-env} we show the variation in the [Fe/H] ratio as a function of distance from the centre of the nearest cluster for the red and complete sub-samples. The median [Fe/H]ratio of galaxies decreases  $\sim 0.05$ dex between the centre and the virial radius (1-2 Mpc) of a typical galaxy cluster at $z\sim0$. Thereafter,
 the [Fe/H] assumes a constant value of $- 0.34$ to upto twice that radius. 
  
 With the general knowledge that galaxies in dense environments are more evolved \citep{blanton05}, it is not surprising to see that on an average the [Fe/H] ratio declines moving away from the cluster centre. Analogous results have previously been obtained for general samples of star-forming galaxies from the SDSS. For instance, \citet{cooper08} not only observed the metallicity-density relation, but showed that it was as strong as the colour-density and luminosity-density relation observed for galaxies in the nearby Universe. These authors also suggest that as much as 15 per cent of the scatter found in the mass-metallicity relation discussed in the previous section can be contributed by differences in the environment. 
 
\begin{figure} 
 \includegraphics[width=8cm]{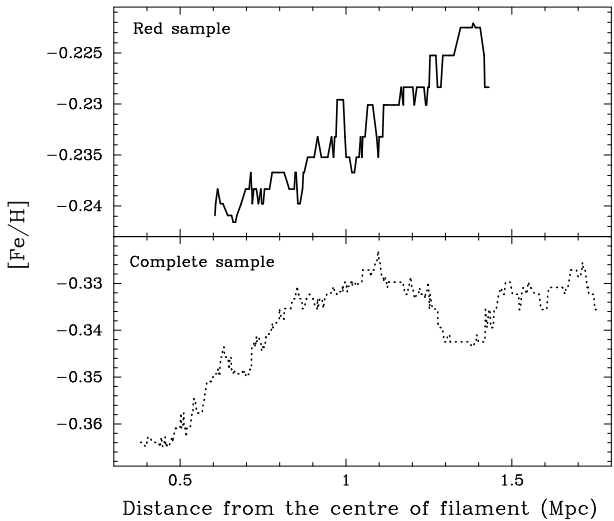}
 \caption{Same as Fig.~\ref{z-env}, but showing median [Fe/H] ratio shown as a function of the distance from the spine of large-scale filaments for the red and complete sub-samples. Assuming filaments to be cylinders, the \Z of galaxies seems to drop towards the central axis of the filament. These observations are in agreement with those
 presented in \citet{SM18}.  }
 \label{z-fil}
 \end{figure}

 Using optical spectroscopic data from the SDSS and UV photometric data from the Galaxy Evolution \\({\it GALEX}) surveys we have shown that broadband colours ({\it g-r, FUV-NUV}) and equivalent width of H$\alpha$ emission line varies as a function of the distance from centre of the filaments in the Coma supercluster \citep{SM18}. These observations imply the critical role played by large-scale cosmic filaments in the evolution of galaxies much before they encounter the hostile cluster-related environmental processes. We now provide further support to these observations in Fig.~\ref{z-fil}. This figure shows the observed trend in the [Fe/H] ratio of filament galaxies as a function of the distance from the spine of the filament, where spine is the central axis of the filament assuming they are uniform cylinders\footnote{The reader is directed to sec.~3 of \citet{SM18} for a detailed description of the procedure used to define filaments. }. The observed trends although mild, suggests that the [Fe/H] ratio is 0.02-0.03 dex lower for galaxies at the spine of the filament relative to their counterparts $\gtrsim 1$ Mpc away from it. 
 
 In line with the literature \citep{alpaslan15, chen15, SM18}, these observations also suggest that galaxies closer to the spine of the filaments either experience greater infall of gas due to the depth of the potential well, or accrete more smaller satellites which aide in lowering the [Fe/H] value. The former scenario seems to be in agreement with the studies of galaxies in galaxy groups \citep{janowiecki17}, where environment is similar to filaments. \citet{janowiecki17} found that low-mass ($M^*/M_\odot \leq 10^{10.2}$) central galaxies in small groups have higher atomic gas fraction, molecular gas content and star formation activity relative to similar galaxies in isolation. In particular, \citet{janowiecki17} found that the atomic gas fraction in low-mass centrals is $\sim 0.3$ dex higher than their counterparts in isolation. These authors attribute their findings to the inflow of cold gas from the cosmic web and gas-rich mergers. 

 Our results are also in broad agreement with the work of \citet{liao19}, who used simulations to show that the gas accreted at high redshift ($z=2.5$, $4$) on the cosmic-web gets shock heated and then cools at the centre of the filaments. These authors find that $\sim 30$ per cent of this gas gets accreted on to the halos on the filaments. Thereafter, this accreted gas will not only lower the metallicity of the halos, but also provide fuel for star formation.

 \subsection{Caveats}
 \label{caveats}
 
In this section we describe some shortcomings of the data and analyses presented in this paper. 
\begin{itemize}

\item{{\it Fibre-spectroscopy}:} In SDSS, galaxy spectra are obtained by observing each galaxy through a $3^{\prime\prime}$ diameter fibre. This means that at the distance of Coma we are essentially receiving light from the central $\sim 1.4$ kpc of a galaxy and the line indices are measured for the core region of the galaxy only. The SDSS Lick indices are not corrected for this aperture bias. Therefore, ages and \Z estimated using these indices, may not be representative of the whole galaxy. The effects of such an aperture bias are described in \citet{gallazzi05}. In the context of the work presented here it is particularly important that even though the age and \Z may be overestimated for some galaxies, the mean trends in age and \Z observed as a function of stellar mass are not significantly affected \citep{gallazzi05}.
 
 \item{{\it Completeness}:} SDSS being a fibre-based survey is limited by the physical size of fibres, such that no two fibres can be closer than $55^{\prime\prime}$ or within $100^{\prime\prime}$ of the plate centre. As a consequence repeated observations of each region on the sky are essential in order to obtain high completeness.  

 \item{{\it Luminosity-weighted age and {\it Z}}:} When interpreting trends in luminosity-weighted, model-based age and \Z it is important to remember that averaging over all stellar populations masks the true complexities in both these quantities. For instance, even a small fraction of very young population together with very old populations can dramatically change the strength of Balmer lines, thus making the galaxy appear much younger than its mass-weighted age \citep[][their fig.~1]{worthey95}.
 
  \item{{\it Sample and data}:} Given the limitations of our sample size, data quality and methodology adopted to estimate the age and metallicity of galaxies, it has not been possible to disentangle the effect of stellar mass and environment completely by refining bin sizes \citep[e.g.][]{peng10}, or splitting our sample into mass bins every where (e.g. Figs.~\ref{z-env} and \ref{z-fil}). Despite these drawbacks, we present our complete analysis in this paper in the hope that future studies on galaxy properties in the cosmic-web will draw inspiration from this work and better constrain the statistical observational trends seen here.
    
 \end{itemize}


\section{Summary}
\label{sum}

 In this paper we have explored the statistical trends in age and stellar \Z of galaxies in the Coma supercluster as a function of their \smass and environment. To do so, we studied the luminosity-weighted mean ages and metallicity of galaxies using the SDSS Lick absorption line indices for two samples - a complete sample of 2,953 galaxies, many of which have large errors in the absorption line indices and a smaller sub-sample of 1,263 galaxies where every absorption line index is subject to an error of less than $20$ per cent. We have also examined the dependence of luminosity-weighted mean ages and metallicity on \smass and large-scale environment of galaxies. Further, we have made an attempt to identify the factor (age, \smass or environment) which governs the metallicity of galaxies, and trends on top of the generic observation that on average, dwarf galaxies have lower metallicity than their massive counterparts. Our key results are as follows:
\begin{itemize}
\item At fixed (luminosity-weighted) age, stellar metallicity of all galaxies, dwarfs and giants, is independent of stellar mass except the most massive ($M^*/M_{\odot} \gtrsim 10^{10.7}$), oldest galaxies ($\gtrsim 9$ Gyr) for which the \Z declines sharply with \smassa. 
\item At fixed (luminosity-weighted) age, metallicity of galaxies declines with age in cluster and filament environments alike. Our complete sample, including galaxies with higher measurement uncertainty suggests that the \Z may increase with \smassa, but no such trend is observed for the low uncertainty sub-sample.
\item Massive galaxies ($M^*/M_{\odot} > 10^{10.5}$) in clusters and filaments have similar distribution for the luminosity-weighted age of galaxies, where as their counterparts in the voids are younger by $\sim 0.5$ Gyr. The age distributions for the intermediate mass galaxies ($10^{10} < M^*/M_{\odot} < 10^{10.5}$) on the other hand show statistically significant differences, such that the galaxies in clusters are older than the filament galaxies by 1-1.5 Gyr, while their counterparts in the voids are younger than filament galaxies by $\sim 1$ Gyr.
 \item The median stellar \Z of filament galaxies declines towards the central spine of the cylinder. This result indicates that some mechanism, most likely the infall of new gas on the cosmic web being accreted by the filament galaxies, is leading to a lower \Z.   
 
 Our observations support a scenario where galaxies are born in the voids, and then assemble into clusters through filaments. During the course of their passage, as these galaxies evolve they will accumulate metals as well as stellar mass, thus explaining the observed mass-density, age-density and metallicity-density correlations, along with the trends observed for the \Z and age of galaxies as a function of their \smass and environment. It would be interesting to explore this hypothesis in simulations and test its impact on the observed properties of galaxies at $z\sim0$. In our knowledge, this is the first study of the age and \Z of galaxies in a supercluster providing a continuous range of environments. If confirmed, the statistical trends observed here would help refine recipes for the evolution of galaxies in different environments in simulations. 
     
 We believe deeper spectroscopic data for a range of environments is critical to confirm the observations made in this work. Although the Coma supercluster provides a unique laboratory where galaxies spanning almost three orders of magnitude in stellar mass could be explored in a continuous range of environments, it may not be a fair representative of the nearby Universe because it is $\sim 3$ times denser than the surrounding space \citep{hogg04, gavazzi10}. Future large facilities such as the Large Synoptic Survey Telescope (LSST), combined with spectroscopic follow-up and 21-cm data will be a key to confirming such trends, and understanding their origin. Large radio surveys planned for the near future may be able to confirm how, and how much of the cold gas is accreted by galaxies from the cosmic-web. 
\end{itemize}

\section{Acknowledgements}
 Funding for SDSS-III has been provided by the Alfred P. Sloan Foundation, the Participating Institutions, the National Science Foundation, and the U.S. Department of Energy Office of Science. The SDSS-III web site is http://www.sdss3.org/.
 
SDSS-III is managed by the Astrophysical Research Consortium for the Participating Institutions of the SDSS-III Collaboration including the University of Arizona, the Brazilian Participation Group, Brookhaven National Laboratory, Carnegie Mellon University, University of Florida, the French Participation Group, the German Participation Group, Harvard University, the Instituto de Astrofisica de Canarias,
the Michigan State/
Notre Dame/JINA Participation Group, Johns Hopkins University, Lawrence Berkeley National Laboratory, Max Planck Institute for Astrophysics, Max Planck Institute for Extraterrestrial Physics, New Mexico State University, New York University, Ohio State University, Pennsylvania State University, University of Portsmouth, Princeton University, the Spanish Participation Group, University of Tokyo, University of Utah, Vanderbilt University, University of Virginia, University of Washington, and Yale University. 

Mahajan acknowledges funding from the DST-SERB INSPIRE Faculty award (DST/INSPIRE/04/2015/002311), Department of Science and Technology (DST), Government of India. 

\bibliographystyle{cas-model2-names}

\bibliography{coma_newastro_arxiv}
\end{document}